# Implementing SCADA Scenarios and Introducing Attacks to Obtain Training Data for Intrusion Detection Methods[1]

Simon Duque Antón, Michael Gundall, Daniel Fraunholz, and Hans Dieter Schotten
German Research Center for Artificial Intelligence, Kaiserslautern, Germany
Simon.Duque_Anton@dfki.de
Michael.Gundall@dfki.de
Daniel.Fraunholz@dfki.de
Hans_Dieter.Schotten@dfki.de

**Abstract:** Cyber-attacks on industrial companies have increased in the last years. The Industrial Internet of Things increases production efficiency, at the cost of an enlarged attack surface. Physical separation of productive networks has fallen prey to the paradigm of interconnectivity, presented by the Industrial Internet of Things. This leads to an increased demand for industrial intrusion detection solutions. There are, however, challenges in implementing industrial intrusion detection. There are hardly any data sets publicly available that can be used to evaluate intrusion detection algorithms. The biggest threat for industrial applications arises from state-sponsored and criminal groups. Often, formerly unknown exploits are employed by these attackers, so-called 0-day exploits. They cannot be discovered with signature-based intrusion detection. Thus, statistical or machine learning based anomaly detection lends itself readily. These methods especially, however, need a large amount of labelled training data. In this work, an exemplary industrial use case with real-world industrial hardware is presented. *Siemens S7* Programmable Logic Controllers are used to control a real world-based control application using the OPC UA protocol: A pump, filling and emptying water tanks. This scenario is used to generate application specific network data. Furthermore, attacks are introduced into this data set. This is done in three ways: First, the normal process is monitored and captured. Common attacks are then synthetically introduced into this data set. Second, malicious behaviour is implemented on the Programmable Logic Controller program and executed live, the traffic is captured as well. Third, malicious behaviour is implemented on the Programmable Logic Controller while still keeping the same output behaviour as in normal operation. An attacker could exploit an application but forge valid sensor output so that no anomaly is detected. Sensors are employed, capturing temperature, sound and flow of water to create data that can be correlated to the network data and used to still detect the attack. All data is labelled, containing the ground truth, meaning all attacks are known and no unknown attacks occur. This makes them perfect for training of anomaly detection algorithms. The data is published to enable security researchers to evaluate intrusion detection solutions.

## 1. Introduction

Malicious actors are taking an increasing interest in manufacturing companies (Duque Anton et al. 2017). Even though attribution in cyber-attacks is always extremely difficult, political interests as well as economic gain are considered to be the main motivators for industrial sabotage and espionage (Rid and Buchanan 2014). The introduction of the Internet of Things (IoT) into the industrial domain, creating the Industrial Internet of Things (IIoT), amplifies the increase in attacks due to an enlarged attack surface. The IoT is known to be vulnerable to cyber-attacks, the revenue created by cyber-crime is higher than most other fields of crime, such as drug and human trafficking, as well as many legit companies (Leyden 2018). This shows the potential in cyber-crime and the motivator for attackers. If enterprises want to participate in the paradigm of IIoT, promising interconnectivity, scalability and flexibility, they have to defend themselves. After three decades of virtually unprotected industrial networks, security solutions are becoming available and relevant. In the beginning of industrialisation, industrial networks were considered secure due to their unique nature and physical separation (Igure et al. 2006) . Many industrial communication protocols do not contain methods for authentication or encryption. The IIoT opens industrial networks to public and insecure networks, increasing the attack surface. As a result, effort has been put into industrial Information Technology (IT) security solutions, such as intrusion detection tools and Security Event and Information Management (SIEM) systems, creating new branches of business. In addition to that, research has shown interest in this field, working on intrusion and anomaly detection systems for industrial applications. One of the largest issues, however, is the generation of training data. In the field of IT security, testing and validating algorithms and tools on sound and realistic data is crucial for its success. For the longest time, home and office IT security solutions were based on signatures and rules, meaning each known attack has a characteristic signature that is used to detect it. Machine learning- and statistics-based intrusion detection has been introduced much later than the first signature-based methods. However, signature-based approaches are not as suited for industrial application as they are for home- and office-based environments. Often, attacks on industrial environments do not exploit software vulnerabilities, but use the systems in the appropriate way, exploiting the design of protocols and systems (Morris and Gao 2013), in contrast to *Stuxnet* for example, that used different known and unknown vulnerabilities (The Langner Group November 2013). If no authentication is used to identify entities, an attacker can easily spoof identities. Furthermore, industrial networks are much more diverse than home and office networks as they are highly application specific. The given characteristics present machine learning and statistical approaches as promising. These methods need a lot of data, ideally containing ground truth, i.e. only known malicious and non-malicious traffic, as well as labels. Such data is difficult to obtain, despite many efforts, there is no standardised data set to benchmark and evaluate intrusion detection data. In this work, we propose a method to inject different types of malicious traffic into Microsoft Object Linking and Embedding (OLE) for Process Control Unified Architecture (OPC UA)-based communication of a real productive system. First, data is collected and then later injected with synthetic malicious traffic. After that, a Programmable Logic Controller (PLC) in the presented setting is infected with malicious software. This malicious behaviour is monitored, together with pre- and succeeding non-malicious behaviour. Finally, additional sensors are introduced to the proposed environment. If an attacker takes control of a PLC, physically separated sensors might be the best chance at detecting the malicious activities. The sensor values are captured and correlated to the network traffic.

---

[1] This is the preprint of a work published in the proceedings of the 2019 International Conference on Cyber Warfare and Security. Please cite as follows:
Duque Anton, S.; Gundall, M.; Fraunholz, D. and Schotten H. D.: "Implementing SCADA Scenarios and Introducing Attacks to Obtain Training Data for Intrusion Detection Methods." In Proceedings of the 2019 International Conference on Cyber Warfare and Security (ICCWS), Academic Conferences International Limited, 2019, pp. 56-65.

The remainder of this work is structured as follows. In Section 2, approaches for generating industrial data sets are introduced. The use case scenario used to generate data is presented in Section 3. After that, the methods to inject attacks are discussed in Section 4. The data generated in the use case scenario in combination with the introduction of attacks is evaluated in Section 5. This work is concluded in Section 6.

## 2. Related Work

There is a vast amount of works related to the detection of intrusions and anomalies in network traffic and host information. Despite the abundance of algorithms and methods, however, few publicly available data sets exist to benchmark and validate the approaches. Especially in the field of machine learning-based anomaly detection, large data sets containing the ground truth, i.e. correctly labelled events, and a sufficient amount of events representing all classes are necessary to evaluate the performance. For home and office network intrusion detection, the KDD Cup '99 data set has been the most famous benchmarking data set (University of California, Irvine 1999). This data set, however, has significant systematic drawbacks that result in a poor evaluation of algorithms. *Tavallaee et al.* present a reviewed version of this data set without these flaws to use instead (Tavallaee et al. 2009). The characteristics of home and office based network data sets are different from the characteristics of industrial ones. As *Morris and Gao* state, most researchers evaluate their algorithms on data sets created in the course of their work (Morris and Gao 2014). In order to provide comparable and sound results, *Morris et al.* present a data set that was taken in a laboratory environment using water and gas storage systems that communicate via *Modbus* (Morris et al. 2011). The drawback of this data set is the lack of some packet information. Instead, they provide already aggregated and pre-processed data. *Lemay and Fernandez* introduce a *Modbus*-based data set containing the communication of a simulated industrial process (Lemay and Fernandez 2016). They also describe their emulation environment (Lemay et al. 2013). *Vasilomanolakis et al.* introduce a generator for realistic intrusion detection data sets that might be adapted to industrial use cases (Vasilomanolakis et al. 2016). *Ring et al.* also propose a generator framework to create synthetic network data (Ring et al. 2017b, 2017a). Their framework consists of scripts to simulate different aspects of network traffic with tuneable hyper parameters. Even though their simulation targets are home and office networks, this concept could be adapted to generate industrial network data as well. *Duque Anton et al.* discuss advantages and disadvantages of different simulation tools in comparison to emulation and real world infrastructure (Duque Anton et al. 2018). Additionally, there is some noteworthy work on testbeds and simulators that allow researchers to generate data on demand. *Siaterlis et al.* propose a testbed that allows analysis of SCADA attacks on industrial networks (Siaterlis et al. 2013). *Genge et al.* propose a similar testbed that allows the generation of industrial scenarios. Real malware can be introduced into these simulations to analyse the effects (Genge et al. 2012).

## 3. Use Case Description

In this section, the preliminaries for the generation of the data sets are introduced. The data sets generated in the course of this work are based on the OPC Unified Architecture (UA) protocol. It is introduced in Subsection 3.1. After that, the process environment that generates the data is presented in Subsection 3.2. It consists of industrial-grade process hardware, enabling it to generate realistic data. The attack scenarios that are implemented and evaluated in this work, are presented in Subsection 3.3.

### 3.1 OPC UA

OPC UA is the platform independent successor of the OPC standard, developed by the OPC Foundation (Technical Report 62541-1). The general aim of OPC UA is the secure, easy and platform independent exchange of information between industrial appliances. Therefore, it provides a communication protocol for the data transport, which is located on the application layer of the Open Systems Interconnection (OSI) model, as well as a full information model. The information model can be used for modelling objects, e.g. single devices or an entire production facility. The loose coupling of information model and transport protocol is one of the major advantages. It provides the possibility to adopt the communication protocol that is required for the specific application. In the majority of cases, the communication on the SCADA level is server-client based. A server provides information about process states, usually obtained from sensors and actuators controlled by a PLC, which can be accessed by clients. A client can be a Human Machine Interface (HMI) or a SCADA system. Since control commands and process alarms or events occur in an acyclic fashion, the loss of one of those data packets would have severe consequences. Therefore, the server-client model of OPC UA makes use of TCP on the transport layer.

There are several implementations of OPC UA. The *Siemens*-modules used in this work provide an implementation for PLC and HMI. Additionally, a *Java* application was used as an HMI that is based on an open source implementation of the OPC UA communication stack.

### 3.2 Process Environment

The data created in this work has been generated with a *Festo Didactic* model, the *Festo Didactic MPS PA Compact Workstation* that has been adapted to the research cases presented in this work. It is shown schematically in Figure 1. It consists of two water containers, 101 and 102, the DC motor of a centrifugal pump M101, a ball valve controlled by a solenoid valve M102. There are four limit value switches, of which two, S111 and S112 are float switches and B113 as well as B114 are capacitive. A PT100 temperature sensor B104 is used, as well as a vane sensor B102, monitoring the flow of liquids from container 101 to 102. A piezo resistive pressure sensor B103 measures pressure, an ultrasound sensor B101 measures the liquid level. All sensor values are submitted as responses to polling and thus contained in the monitored data. The process allows setting an upper and lower threshold for the water level in one of the containers. If the water level falls below the lower threshold, the pump is activated and refills the water in the given container. If the water level rises above the upper threshold, e.g. after the thresholds have been changed, the ball valve is opened to release the excess water. Information about the water levels, temperature and pump activity is polled periodically by HMI and *Java* application respectively. The process is exemplary for industrial batch controls (International Standard 61512-1). It is offered by *Festo Didactic* as a training environment and displays realistic behaviour.

### 3.3 Attack Scenarios

Security is a major aspect in the OPC UA Standard. Therefore, the communication protocols that are part of the standard provide security mechanisms, such as encryption. In contrast to many older industrial communication protocols, such as *Modbus* and *PROFINET* (International Standard 61784-1:2014), authentication and encryption are available. A report of the *German Federal Office for Information Security* from 2016 concludes the mechanisms in the OPC UA protocol is reasonably secure in *SignAndEncrypt* mode (Damm et al. 2016). In the security mode *Sign*, eavesdropping is possible, violating the security objective of confidentiality. If the security mode *None* is employed, spoofing and alteration of messages is possible as well. Furthermore, the protocol cannot prevent message flooding and server profiling attacks. However, flaws in the different implementation versions were discovered. *Kaspersky* discovered 17 vulnerabilities due to implementation errors in OPC products (Targett 2018). In this work, no implementation or protocol flaws are exploited or researched. Instead, the attack scenarios rely on an attacker having gained access to a PLC and exploiting the network further. The attack vector leading to the attacker's access on the PLC is not part of the consideration. In most cases, the human factor presents the gravest threat (Gonzalez and Sawicka 2002), enabling attackers to break the perimeter. This paper only focusses on possible actions of an attacker after breaking into a network. To do so, three attack scenarios are introduced. Overall, three data sets are created. The data sets differ in the way they are created. In each data set, one or two attacks are introduced. The attack scenarios are based on the work of *Morris and Gao* who introduced the different kinds of attacks on *Modbus* applications (Morris and Gao 2013). Response and measurement injection as well as command injection attacks have been summarised as attack scenario 2 in this work. Furthermore, Denial of Service (DoS) attacks have been omitted.

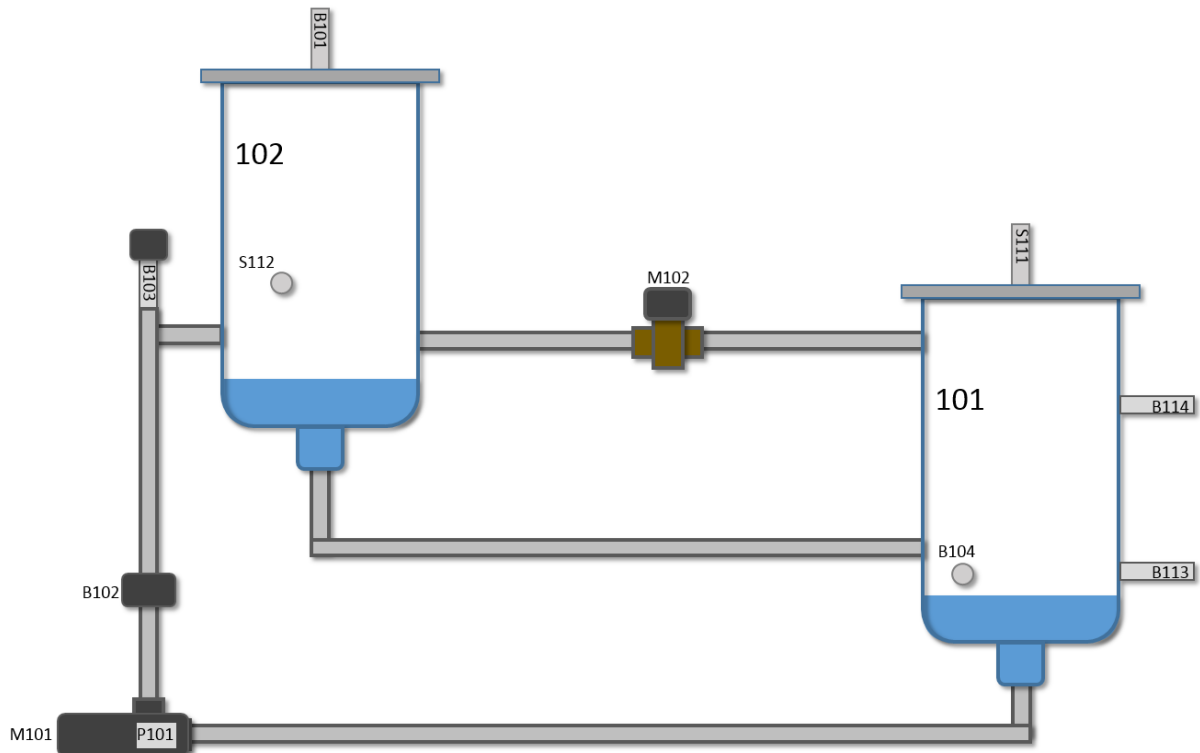

**Figure 1: Schematic of Process Environment**

### 3.3.1 Scenario 1

The first attack scenario is an attacker attempting reconnaissance in the network by enumerating services and devices. Scanning is the second phase of an attack according to classic taxonomies (EC-Council 2011). Attackers evaluate the environment in order to find suitable targets. If an attacker gained access to an industrial network, vulnerable devices need to be found before the actual attack can take place. Additionally, scanning can often be used to detect and mitigate attacks before they take place, as it changes the network traffic characteristics (Lazarevic et al. 2003).

### 3.3.2 Scenario 2

The second scenario is an attacker trying to hinder operation by introducing malicious values into the responses to the HMI. This is called response and measurement injection as well as command injection attacks by *Morris and Gao* (2013). The aim of an attacker in this scenario is to stall a process by introducing malicious behaviour and provoking maintenance actions. In industry, availability is considered the highest security objective of the CIAA tetrad that has been extended from the well-known CIA triad (Andress and Rogers 2011). CIAA describes the classic IT security objectives Confidentiality, Integrity, Authenticity and Availability. Stalling production and provoking maintenance will delay production and make the operator lose money.

### 3.3.3 Scenario 3

The third scenario describes an attacker wanting to damage the application. The values responded to the polling mimic correct behaviour, but the actual machine behaviour deviates. In contrast to the attack described in Scenario 2, this kind of attack leads to faulty process results. In process automation, this can lead to faulty products that do not meet their specifications. This kind of error can be fatal for industry, as the faulty products will lead to great financial loss as the fault will be discovered with a delay. Another use case for this attack scenario is the sabotage of production facilities. Operating devices outside their specification while providing feedback that the operation is working as intended leads to an increase in wearing of devices. Furthermore, damages can occur due to wrongful operation that will produce loss of production, damage to machines and possibly damage to further entities or human operators. The *Stuxnet* attack aimed at damaging centrifuges by operating them outside their specification (The Langner Group November 2013).

## 4. Attack Injection Approaches

In this work, two different approaches to inject attacks into the data that was generated and monitored were employed. This was done for two reasons: First, the variety of attacks and their characteristics is supposed to be as wide as possible. And second, employing different techniques of attack injection allows to draw conclusions about the feasibility and soundness of simulation approaches. If the characteristics of synthetic and real attacks are similar, data generation can be performed with reduced effort. A special focus lies on the question which characteristics can be generated synthetically and which cannot. In this section, the two approaches are introduced.

### 4.1 Approach 1: Synthetic Attacks

For the first approach, synthetic attacks are introduced into data sets of real world devices. The attacks are created with *Scapy*, a *Python*-based packet manipulation program (Biondi 2018). It can be used to read the packet capture and create new packets that only differ in certain fields. This approach was used to change the payload of data set 1. The advantage of this approach is the simplicity, as malicious packets can be introduced to any traffic capture. The downside, however, is the synthetic origin of attacks and the challenge that some attacks might contain features not present in synthetically created packets.

### 4.2 Approach 2: Real-World Attacks

In the second approach, the application hardware is programmed in a way an attacker could do in order to achieve the malicious intent. No post processing of data is required, all packets are provided as they were monitored in the network. This provides the highest possible degree of realism. The attack scenarios, however, have to be chosen in a way that resembles realistic attacks.

## 5. Evaluation

In this section, the data sets that were created in the course of this work are presented and evaluated. In total, three data sets were created. They consist of several files each. An overview about their characteristics is provided in Table 1.

Table 1: Data Sets and Their Features

| FEATURES | DATA SET 1 | DATA SET 2 | DATA SET 3 |
|---|---|---|---|
| **DURATION** | 41 minutes | 41 minutes | 10 minutes |
| **ATTACK INJECTION** | Approach 1 | Approach 2 | Approach 2 |
| **ATTACKS** | 2 | 2 | 2 |
| **NUMBER OF FILES** | 5 | 5 | 4 |

The data sets are available at the following link for analysis and evaluation: https://projects.dfki.uni-kl.de/IUNO/

## 5.1 Data Set 1

The first data set contains the communication of a PLC with a *Siemens TP700* HMI as well as a web frontend application. The communication with each kind of HMI is split into two separate files. Two entities participate in the TCP-based OPC UA communication, IP addresses 192.168.5.21 and 192.168.5.12 respectively, corresponding to the HMI and 192.168.5.51 denoting the PLC. The trace contains 41 minutes of operation during which close to 5000 OPC UA packets were transmitted. These packets contain the values of sensors and actuators, as shown in Figure 2. They contain regular behaviour, as can be expected from periodical processes. With the help of *Scapy*, two attacks are introduced. The first one sets the all values to 0 in OPC UA packets 1500 to 1700. The second attack mimics the real behaviour, but with half the frequency. This can be seen in OPC UA packets 3000 to 3500. Both attacks are shown in Figure 3.

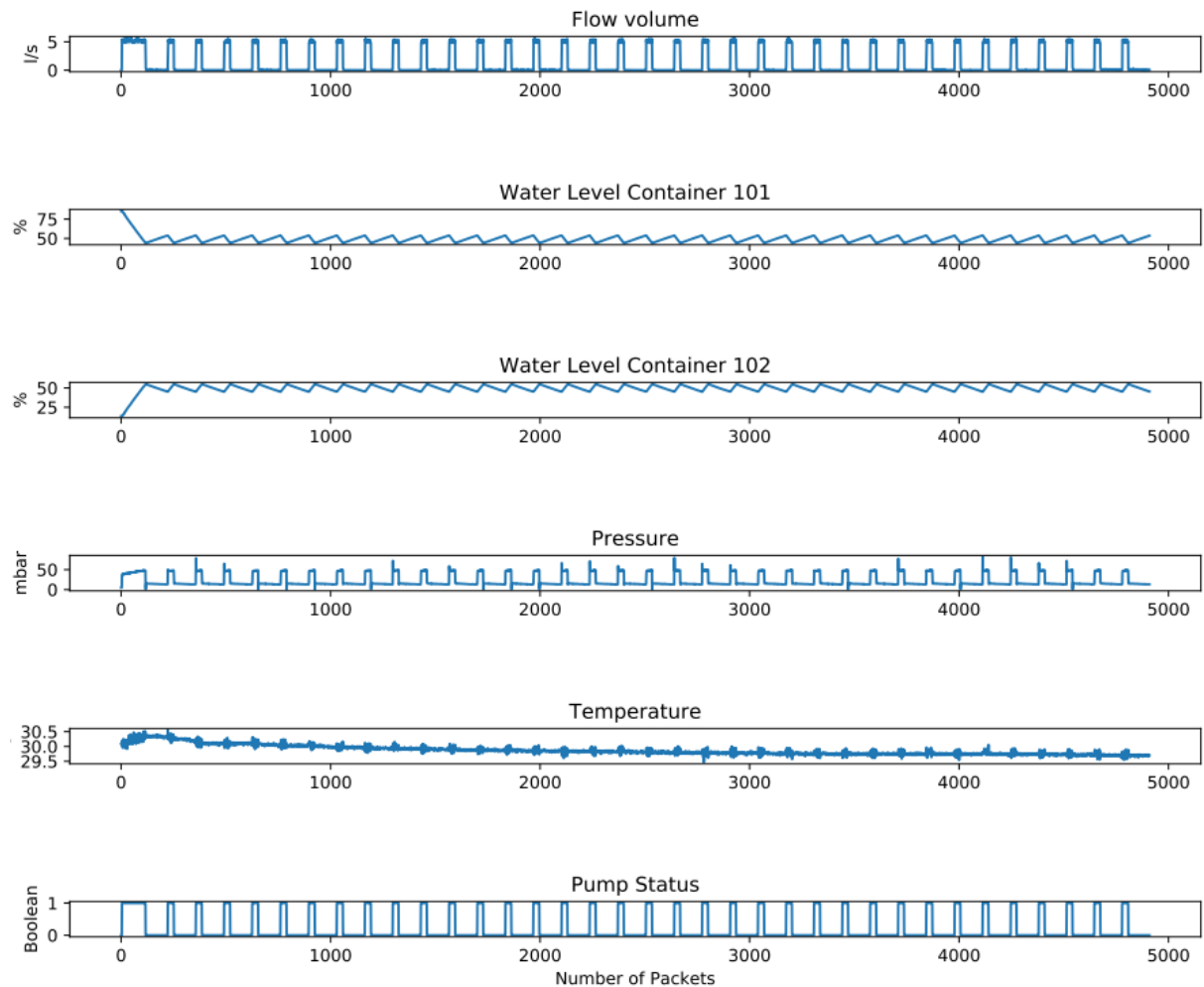

**Figure 2: Normal Behaviour in Scenario 1**

## 5.2 Data Set 2

In this data set, two different attacks are implemented. First, the address range is scanned by a malicious OPC UA client, implemented as a custom, *Java*-based HMI. After the PLC is discovered, the values of the pump condition are inverted. Second, wrong sensor values are returned by the PLC. In contrast to data set 1, this is done by the PLC itself. The malicious values look like in data set 1. In addition to the falsified values, a high amount of Address Resolution Protocol (ARP) traffic is observed that comes from an enumeration of the address space by connecting to incremental Internet Protocol (IP) addresses.

## 5.3 Data Set 3

For the third data set, side channel sensors were employed. The PLC was configured to provide sensor values as if the operation was working correctly. This leads to network traffic as observed in data set 1 without malicious traffic. If an attacker has access to a PLC as such, this behaviour would allow the attacker to damage the production process without operators noticing from HMI information. The side channel sensors employed included a microphone, a temperature and a flow sensor. As a temperature and flow sensor, the internal sensor was used, since it provided the required values easily. However, external flow sensors are readily available. Two attacks were implemented: First, the pump is not shut off, despite the predefined water level being reached. This results in the pump running empty. Second, the ball valve control is set to one constantly, resulting in a stream of water leaking from container 102.

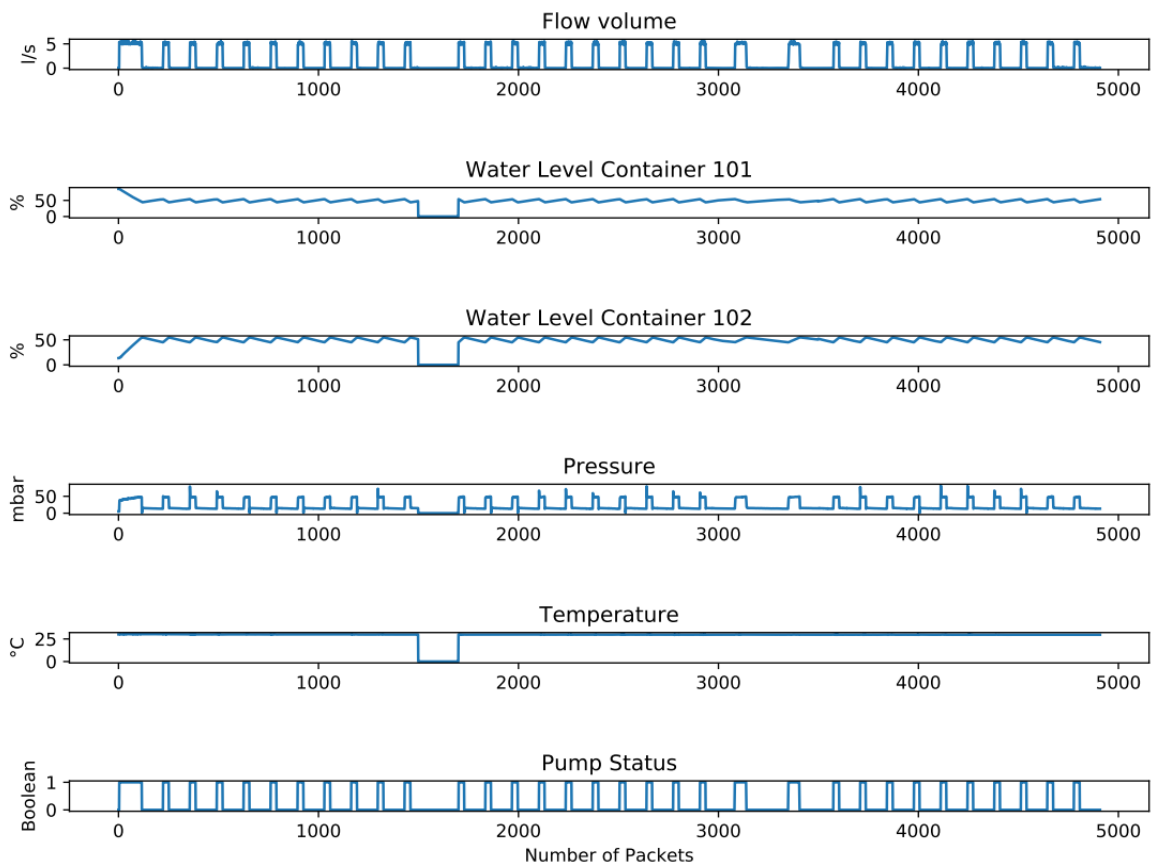

Figure 3: Attacks in Scenario 1

The microphone provides the most promising data. The spectrum of normal and malicious pump operation are shown in Figure 4. The normal operation is drawn in blue, the malicious one in orange. There are distinct peaks between 500 and 600 Hz, as well as at 1000 Hz. The first attack scenario is distinct from this behaviour, as a peak at 300 Hz can be observed that is higher than the normal peaks in Figure 4. Furthermore, the overall spectrum is higher, the peak value is -10 dB. The second scenario can be detected as well, as the constant stream of water leads to a different spectrum than the spectrum if no activity happens.

The flow is zero, as no water is left in container 101 to be pumped into 102. This is shown in Figure 5. Thus, flow can be used to distinguish normal from anomalous behaviour. Furthermore, the temperature value shows more short-term changes than if no attack is taking place. However, the absolute range is below one degree Celsius. The flow value is also capable of detecting the second attack, as the intervals during which water is pumped from 101 to 102 are longer than the previous ones. The difficulty lies in determining the difference between a change in process and an attack.

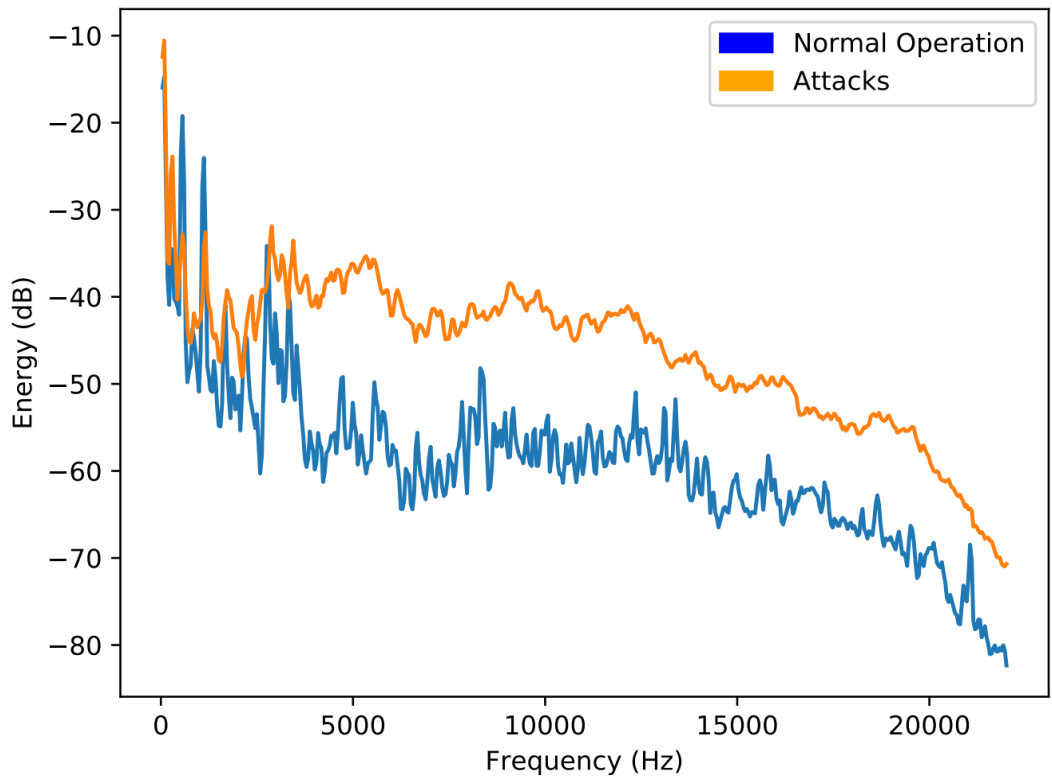

**Figure 4: Normal and Malicious Pump Operation Spectrum**

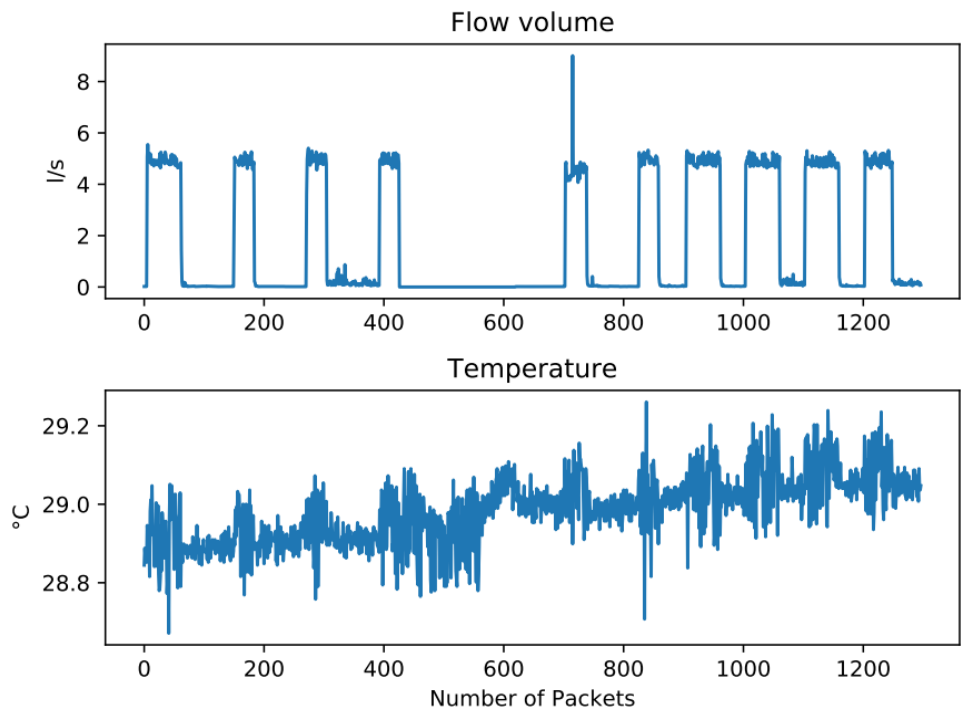

**Figure 5: Flow and Temperature in Data Set 3**

## 6. Conclusion and Outlook

In this work, three data sets were generated with two different approaches. These data sets contain the operational traffic of an industrial process to which attacks have been introduced. The attacks differ in their structure and effect of the environment. Additionally, the attacks differ in characteristics that make them recognisable. This enables researchers to develop and evaluate methods for industrial intrusion detection. The presented scenarios are exemplary for threats to industrial environments. Furthermore, this work provides real-world examples that can be taken as a template for simulation environments. There are many possible approaches to detect anomalies, such as timing- or packet-based, and depending on the approach, a different feature set is required. These features can be simulated in accordance with the provided data, while omitting non-required features.

As a next step, a simulation environment is developed to provide data of arbitrary size where arbitrary attacks can be introduced. Furthermore, the introduction of context, such as Enterprise Resource Planning (ERP) and Manufacturing Execution Systems (MESs) provide, is planned. Context as a feature for anomaly detection shown promise. Additionally, Multicast Domain Name System (mDNS) is to be employed for enumeration, as there are implementations for OPC UA available.

## 7. Acknowledgements

This work has been supported by the Federal Ministry of Education and Research (BMBF) of the Federal Republic of Germany within the project IUNO (KIS4ITS0001). The authors alone are responsible for the content of the paper.